# An Open-Source Monitoring System for Remote Solar Power Applications

Nikolas Wolfe
CIS 4914 Senior Project
Advisor: Dave Small (dts@cise.ufl.edu)
12/11/09

# Abstract

Renewable energy systems are an increasingly popular way to generate electricity. As with any new technological paradigm, new challenges have emerged which are unique to the utilization of renewable energy systems. One of these challenges in particular is the development of effective monitoring technologies to compensate for the decentralized nature of remote power generation. This project details the development of an open-source monitoring system for remote solar power systems. The problem space that this project is specifically concerned with deals with the reduction of cost and the use of open platforms to make solar monitoring viable in developing countries where both the resources and general knowledge required to undertake such efforts are particularly scarce.

Currently, solar monitoring technologies are expensive, limited in their application, and for the most part proprietary. It is arguable that such systems can be developed using non-customized hardware and open-source software that can be obtained and run anywhere in the world. This project is one such argument. This proof of concept is sufficient to show that solar remote monitoring is neither expensive nor particularly cumbersome to implement and thus warrants further investigation and development by the open source community.

# 1   Introduction

## 1.1 Problem Statement

The ability to remotely monitor data about the power output of solar panels and the state of battery banks is of critical importance to the proper long-term maintenance of solar energy systems. Unfortunately, existing remote monitoring technologies are expensive, limited in their application, and in general require paying service fees to third parties above and beyond the cost of basic communication. In places where solar energy systems are employed as a replacement for a lack of electrical grid infrastructure, this cost may lead to the abandonment of the idea of remote monitoring altogether. Thus, the reality of expensive and proprietary remote monitoring technologies holds hostage the viability of renewable energy systems in developing countries.

## 1.2 How Remote Monitoring Works

Solar power systems tend to exist in two flavors–remote and grid-tied. Remote systems are not connected to the electrical grid and serve as a primary power source. Grid-tied systems are integrated with the electrical grid and tend to serve as an auxiliary power source. The system proposed and implemented here is developed with a remote system in mind. Remote solar power systems use devices called charge controllers to apply charging algorithms to banks of deep-cycle batteries. Controllers are essential because using solar panels to charge batteries is not a trivial task; a delicate balance must be struck between the need for batteries to be charged using well-defined and consistent charging cycles and the fact that the output of solar panels can be inconsistent and erratic depending on the weather.

Monitoring is conducted by interrogating a charge controller for data about the performance of a solar power system, and transmitting that data to a remote location or specific person. A cellular modem or other transmission device is generally the vehicle for this transmission. Many solar controllers have serial ports built into them that can be polled for data about a given system using some transmission protocol. A useful diagram to understand the high-level process of remote monitoring is shown below.

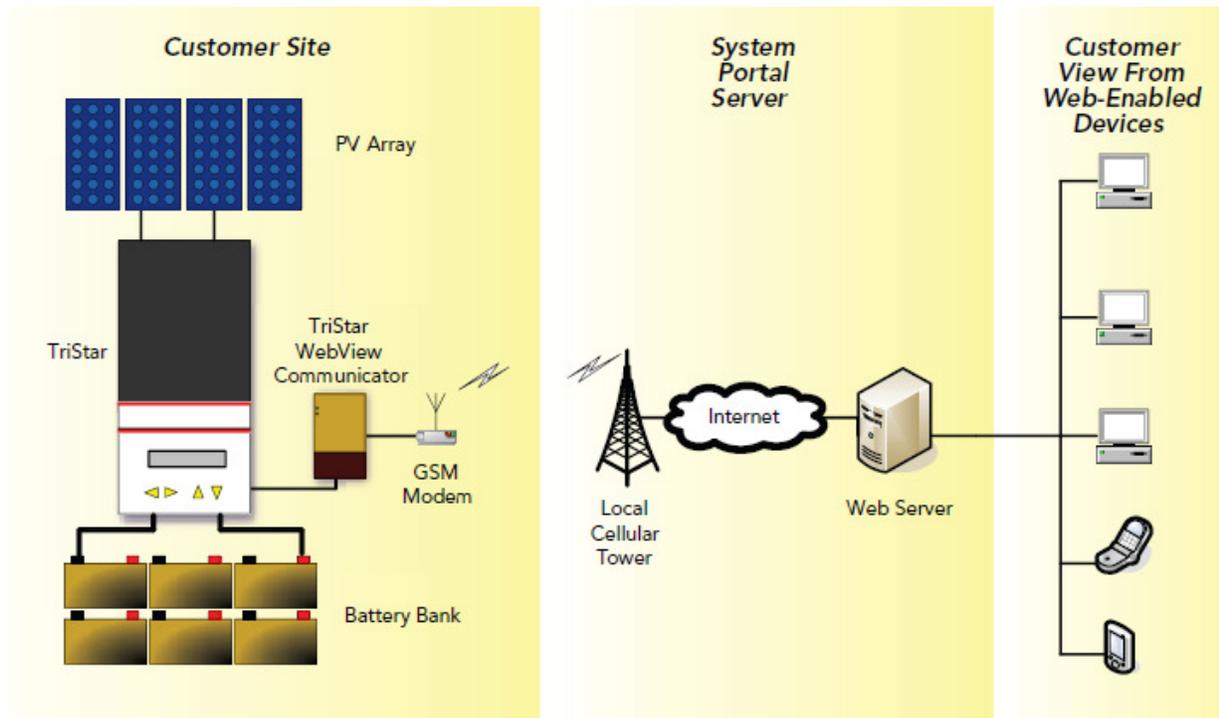

Fig 1: The Morningstar TriStar Web View proprietary remote monitoring system

*1.3 How This Project Departs From Existing Monitoring Systems*
This project represents a significant departure from existing solar monitoring technologies because it is specifically intended to be open-source in every aspect outside of the actual solar power system itself. Furthermore, this problem has never before been phrased in terms of its application in the context of international development or developing countries.

*1.4 Solution Statement*

The purpose of this project is to design and implement an open-source monitoring system for remote solar energy power systems that can deliver useful diagnostic information to system overseers. This system is divided into two different parts. The first is a hardware-centric component that interfaces with a solar charge controller to determine diagnostic information about a power system. This component will process obtained data and transmit it via a telephony communication protocol to a server or possibly a specific person. The second component of the system is the recipient software on the other end of this transfer. This software shall store historical data about the solar system and provide a web interface through which long term statistics can be determined and effective monitoring can be conducted.

*1.5 Contribution*

This project offers the first stab in the direction of a serious and robust open-source monitoring system that can be used and maintained. There are many applications for a system like this. Monitoring a solar power array is just one of the possibilities. Practically any device with measurable outputs running in a remote environment represents a potential future extension of this project. Furthermore, while I do not have the time or the space to fully do justice to the claim that this will be an asset in international development projects involving the use of solar technologies, I maintain that the cost of remote monitoring is impractical in developing countries and thus an open-source system represents a one possible way to make the essential exercises of proper maintenance possible.

## 2   Problem Domain

In general, remote monitoring is a useful idea. As the world's energy economy gradually evolves towards greener and more renewable sources of energy, power generation will become more decentralized. In contrast to the current system of large, centralized power plants servicing vast regions of energy consumers, the energy infrastructure of the future will see many different forms of power generation taking root wherever the demand exists. Coordinating such a vast system will prove impossible without a robust monitoring infrastructure. The ability to measure spot and regional performance is of critical importance if the power of decentralized generation is to be harnessed and the grid is to in fact become smarter. In essence, remote monitoring will be the glue that ties the intelligent grid of the future together.

Having stated the general context of the application domain, I must again return to how this project differs from the usual paradigm. This project is specified with the most meager economic means in mind. It uses only widely available and cheap technologies and explores the non-conventional use of, for instance, an old cellular phone as a data transmission device. As far as rules of thumb are concerned, every design consideration in this project has been made with respect to the likelihood that such efforts could be repeated *anywhere* in the world.

## 3   Literature Review

There is not much literature in existence that explores the specific concept of remote monitoring as an open-source problem or one that should be considered in the context of international development. However, literature on the core concepts of remote monitoring and existing implementations can be found as well as information on the technological realities of operating in developing countries.

*3.1 Existing Monitoring Systems*

The industry leading monitoring system is the Sunny Webbox, a remote monitoring device maintained by SMA Solar Technology, a German solar energy equipment supplier. The Sunny Webbox is essentially a glorified modem that plugs into an inverter and allows system overseers to remotely monitor the output of a given solar power system over the Internet. Unfortunately, the communication protocol between the Sunny Webbox and an actual SMA inverter is not well documented or supported by SMA so it is difficult to program against. Furthermore, the Sunny Webbox requires wired Ethernet in order to remotely transmit data, which is an obviously crippling dependency if we intend to monitor devices operating in truly remote environments or in places without adequate internet infrastructure.

Another industry-leading system is implemented by Fat Spaniel, a company that provides high quality web-based solar monitoring for a monthly fee. Fat Spaniel employs cellular modems to remotely transmit data from a given solar controller. One solar energy company that offers a monitoring service (depicted in Fig.1) through Fat Spaniel is Morningstar Solar. Morningstar Solar sets itself apart from SMA in one particular respect because they use the open Modicon Modbus transmission protocol to allow external devices to interrogate their controllers. They provide the specification for their implementation of the protocol and thus allow other implementing technologies to be used with their controllers relatively easily.

*3.2 Internet Resources and Hardware Choices*

The solar controller that I designed this project to be compatible with (at least as a proof of concept) is the Morningstar TriStar-45 solar controller. This is on account of several factors, including my personal experiences working with Morningstar solar controllers in the past, the availability Morningstar controllers internationally, and the fact that applications can be easily developed to work with Morningstar controllers because they use the open Modbus hardware communication protocol.

The Modicon Modbus protocol reference guide is freely available online, and also includes sample code for more tedious implementation details such as the creation of a working Cyclic Redundancy Check (CRC) function. I used this specification to implement the software required to interrogate the TriStar-45 controller. Morningstar offers all of the required documentation to develop applications in conjunction with their controllers online. The "TriStar Applications Guide", for instance, describes sample configurations for data acquisition and remote monitoring of solar power systems.

The other major technology that is utilized in this project is the Arduino electronics prototyping platform. Arduino is an open-source hardware platform intended mostly for hobbyists and casual hardware enthusiasts. All hardware part listings, board schematics, and documentation required to build Arduino clones can be found and downloaded online. The software environment required to write and upload programs to Arduino boards is also freely available and completely open-source. The Arduino hardware environment is an ideal platform for a project like this because it allows the required hardware to be reproduced in any context where the parts can be found. In an effort to demonstrate this possibility, I have opted to use a Freeduino board instead of an Arduino board. Freeduino is an open-source clone of the Arduino Decimilia board, and can be purchased from hobbyist electronics websites as an unassembled bag of parts and a cut PCB board for a fraction of the price of a preassembled Arduino board.

The last piece of hardware technology utilized in this project is a Motorola W260g cellular phone, which I have engineered into a data transmission device. I have opted to use this phone because of its price, but I would argue that no part of the hardware design in this project is dependent on the specific design of the Motorola W260g. The ways in which these devices interact and accomplish the task of solar remote monitoring is discussed in further detail in the Solution section. It would be a tedious and exhausting task to enumerate the complete set of online resources and documentation available for the products listed above, so I will suffice it to refer the reader to the References section of this report.

*3.3 Academic Background and Personal Experience*

It would have been impossible to come up with the idea for this project without significant personal experiences prior to this point. I have studied solar power systems independently and have a certification in photovoltaic design and installation from Solar Energy International. Furthermore, the inspiration for this project came as a result of an international project that I myself worked on. During the spring and summer of 2007 I worked as an engineer on a school-building project in The Gambia, which employed solar technology as a way to generate electricity and pump water. My job was to design and implement the solar power system for both of these applications, and impart the knowledge required to maintain them to local project stakeholders. While an exhaustive discussion of the successes and failures of this project is well beyond the scope of this paper, I will suffice it to say that during the course of this project that the issue of remote monitoring became something of significant concern.

The semester prior to my departure to The Gambia was spent studying literature on the experiences of various international development organizations that employed solar technology in their projects. This study revealed that the vast majority of such projects fail after their initial completion because of lack of follow-up and failures to incorporate local communities into the maintenance process in any significant way. One thing in particular that is almost always present in accounts of the failure of solar power systems in rural and poor communities is the lack of resources and knowledge required to maintain them.

If we consider the components required for the effective maintenance of a solar power system, the problem begins to take on the shape of an information technology problem. The essential challenge is the proper cycling of batteries based on available sunlight and expected energy demands. All that is truly required to do this is the ability to track system state. Thus, any technologies that can obtain and analyze performance data are of enormous utility. This is of course the fundamental task of remote monitoring technologies, so it is a fairly straightforward case to make that the availability of such technologies in places where solar power systems are implemented can have a significant impact on the ease of maintaining those systems.

## 4   Solution

If we accept the argument that maintaining a solar power system is essentially a challenge of obtaining critical information about the system and taking appropriate action, we can propose a solution in the form of a remote monitoring system. However, there are more concerns than just this. One of the greatest problems with remote monitoring, again, is its cost. So, a system that is intended to serve the purpose of remote monitoring in developing countries and rural environments must be inexpensive in addition to being effective. The most important factor to minimize is the cost after installation.

*4.1 High Level Overview of the System*

This system is broken into two parts. The first is a hardware device that interfaces with the solar controller, polls it for information, processes that information into a format that can be transmitted via some telephony communication protocol (in this case DTMF) and then transmits that data to a remote server or potentially a specific person in the case of a problem. The second part of this system is a software program that will collect the data transmitted, store it for analysis and historical purposes, and allow diagnostic information to be easily displayed via a web interface. I was not able to complete this portion of the system so I will not discuss it in any depth here.

The hardware system that this project uses is designed with certain constraints in mind. First, the system must not rely on customized hardware. I do not consider the solar controller to be a part of this specification because it can operate independently of any monitoring activity and is a general requirement of a functional solar power system. In the case of the microprocessor that I am using to poll the controller and translate that data into DTMF tones, I have not written any code with a proprietary operating system or compiler in mind.

The next major constraint is that the fundamental logic of this hardware system must not depend on a specific solar controller communication protocol. I have completely abstracted this aspect of the system into a simple C++ object that uses a generic interface that betrays no detail of any underlying hardware protocol. This interface will have to be re-implemented for every solar controller that has a different hardware protocol than the Morningstar Modbus implementation, but this is of course to be expected.

The last major constraint of this project is that it is entirely open-source. Nothing that is implemented in this project utilizes any proprietary software, and no hardware is being utilized for a purpose that requires permission or specific royalties to a given company. The cellular phone that I am using is again the Motorola W260g, and its circuitry *is* actually a proprietary design, however I am not using any function that is not generic to every cellular phone, so this fact is rendered irrelevant.

*4.2 How the Parts Fit Together*

The Freeduino design that I am using is called a MaxSerial board, which has an RS-232 port built in. This is required to interface with the Morningstar Solar controller, which allows external devices to interrogate it through an RS-232 9-pin serial port. The Freeduino board is based on the Arduino Decimilia specification, which uses an ATMega328 16 MHz microprocessor, and has 14 digital I/O pins. I use these pins to operate a circuit that is soldered onto the keypad of the Motorola W260g cellular phone. There are fourteen inputs on the cellular keypad which are of concern to this project and they are common to all cellular phones: The first twelve are the keys of the cellular DTMF keypad and the last two are the power and start buttons.

In order to eliminate the tedious process of recharging the battery of the cell phone independently of powering the Freeduino board, the positive battery terminal is broken out with a wire and connected to the +5V output on the Freeduino board. This allows for the keyboard of the cell phone to be powered without the battery, and eliminates a source of potential complication. The last circuit which must be built onto the cellular motherboard is an interrupt-circuit for incoming calls.

Optoisolators are used in many places in the circuitry that manipulates the cellular keypad. Because the mechanical action of a finger pressing a key is represents an electrically isolated event on the keypad, in order to properly simulate this function with control signals from a microprocessor, each key-press circuit must be electrically isolated. Once the ability to press keys on the keypad is abstracted, the data obtained by polling the solar controller can be converted to DTMF tones by simply pressing the right keys in order. The keypad offers the values of 0-9, which is of course a decimal system. The values of the voltage and current of the panels and the batteries from the solar controller can be simply pressed verbatim into the keypad when a call is connected. To allay confusion for the software system that must interpret this string of DTMF tones on the receiving end of call, the '#' key is utilized as a field separator.

## 5 Results

The hardware side of this project, which represents the much more problematic and complicated aspect of this project, was accomplished successfully. The software side was not. In this section I will describe the implementation of the data transmission device, and the software specification is discussed in further detail in the Conclusion section.

The first hurdle of this project was the creation of the appropriate circuitry to interface with the cellular phone motherboard. I was able to use a combination of careful soldering and some creativity with fork-terminals to build the appropriate circuitry to break out the nodes on the keypad for external manipulation. The next great hurdle was figuring out how to switch the keys electronically. My first approach was to use tri-state buffers. Unfortunately, the control signal is not electrically isolated from the load to be switched. Optoisolators ended up being a ready and cheap solution to this problem.

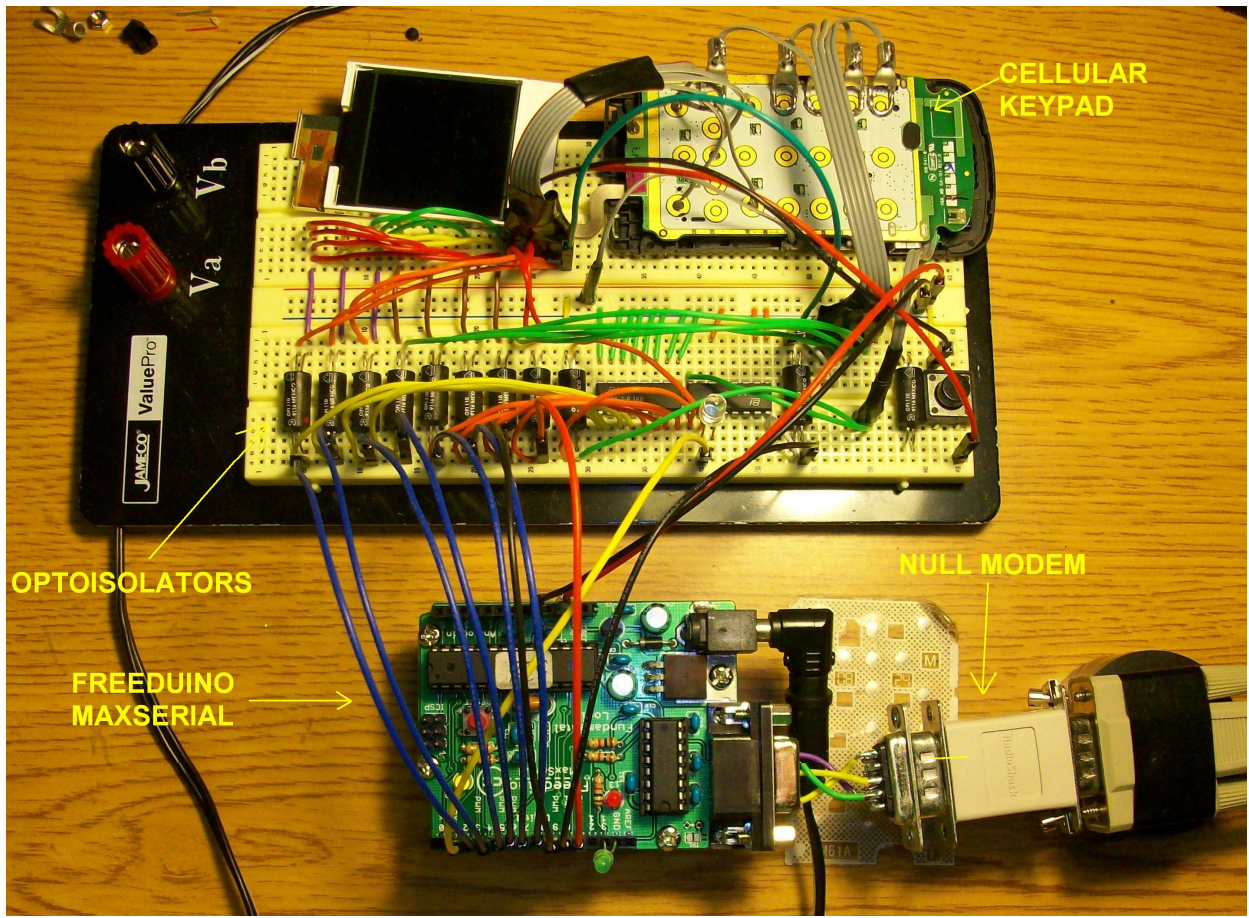

Fig 2: Completed Hardware Implementation

*5.1 Hardware Code*

The Arduino compiler requires that two functions be implemented for any program to be uploaded to a board. These are fairly intuitive and straightforward: there is a `setup` method and a `loop` method. The function of `setup` in this case is straightforward with one exception. After all other initializations have been completed the `setup` function then asserts the power button on the cell phone for a period of 4 seconds, and then waits for 20 seconds. This is the action required to turn on and allow the cell phone to reach a state where it is ready to receive calls. Once inside the main control loop, the logic of the program is also fairly straightforward.

```
void loop()
{
  // check if time to do a transfer
  if( incoming call interrupt has fired )
  {
    // output our data frame
    start the call;
```

```
      output current data;
      terminate the call;
   }

   // poll for data every 30 seconds
   if( 30 seconds has not elapsed )
   {
       wait briefly;
   }
   else if( 30 seconds has elapsed )
   {
      // update the data fields
      log and queue old data;
      poll the controller for new data;
      validate polled data;
   }
}
```

Fig 3: Main Control Loop

An externally fired asynchronous interrupt will set a flag that is checked during each iteration of the loop until it is set, and when this is the case then we know a call from the remote server has been initiated. The call is connected, and a queue of data is output to the cell phone. The rest of the loop is concerned with updating the current data values. This is not implemented as a timed interrupt because these two processes should happen synchronously. Once one of these processes has begun, it must proceed until finished.

The logic for the updating of the system data is the only slightly interesting part of this code snippet. When the controller is polled, the data received is then validated. If the voltage of the battery bank is found to be lower than a pre-set cutoff threshold, it is assumed that the bank is being overdrawn and an alarm is started. This alarm occurs in the form of a call to a specific person. This alarm has the potential to be thrown on every update, so if there is a problem, the system overseer will be alerted continuously until appropriate action is taken. (Appendix A contains a diagram that shows the recorded data from a day-long test and shows the point at which the low-voltage cutoff alarm was fired while the system was under load.)

The other interesting part of this code is the logging and queuing functionality. This is intended as a cost-saving mechanism. Every time a call is initiated, a minute of paid time is used up by the cell phone (even if the call is less than a minute). Thus it behooves us to queue updates and retrieve them in batches that can take up to but never exceed one minute of time to transfer. Since the maximum number of data frames that can be guaranteed to fit in the span of a minute has been determined to be no more than 3, the queue is made of only 2 previously stored data frames, plus the most recently polled data (which may or may not be queued yet).

*5.2 Modbus Protocol Implementation*

The implementation of the controller transfer protocol is a different software module entirely. In order to accomplish this I created an Arduino software library called `ControllerTransferProtocol`. This library allows an object instance to be asked for several items of information, including such functions as `getBatteryVoltage`, `getPwrSrcVoltage`, `getChargeCurrent`, `getLoadCurrent`, and `getTotalKilowattHrs`. The way that any one of these functions works is the same. The Morningstar TriStar-45 controller can be polled using the Modbus protocol.

There are two flavors of Modbus, the ASCII specification and the RTU (Remote Terminal Unit) specification. The TriStar-45 uses the RTU specification of Modbus. There are several functions in the Modbus protocol, but the only one that this project is concerned with is called the "Read Holding Registers" function, and it is named using the binary equivalent of hexadecimal 0x03. In order to call the Read Holding Registers function, we need the address of the device we want to poll, (when there is only one device to communicate with, the address of the TriStar-45 is 0x01), the code of the command (0x03), the address of the starting register we want to read, the number of registers to output consecutively after this register, and then a 16-bit Cyclic Redundancy Check which is used to test that the message was received intact. In the case of a different controller with a different protocol, the result of a call to `getBatteryVoltage` would be the same to the calling logic, so the implementation of the transfer protocol is hidden from the main logic of the program that translates this value into a form that can be transmitted over the air waves as a string of DTMF tones. (See Appendix C)

## 6   Conclusion

This project has some significant claims to success as well as some admitted failures. The greatest success was the completion of the hardware component of this project. The most obvious failure is the fact that I was not able to complete the server-side software that polls and stores data obtained from the solar controller. However, the accomplishment of the hardware side of this project represents the critical proof of concept for this system because not only is the hardware the most significant initial cost in any remote monitoring system, it also represents the source of most after-installation costs. Thus, the ability to show that this implementation accomplishes the requirements of this project is much more contingent on the success of the hardware side than the software side. It is conceivable that a simplified gradient of this project could actually operate independent of this logging software altogether.

*6.1 Lessons Learned*

The lessons of this project are manifold. To start, I learned a great deal about the pitfalls and travails of working with hardware. While I was already aware of most of these things, the fact that I was designing and implementing something that had never been done before was both an invigorating and humbling intellectual exercise. In the past, when I have done assignments for classes, projects are always approached with the overriding reassurance that a solution to a given task exists and has been accomplished before. In this case, there was no such reassurance. The difficulty of dealing with problems was of exponentially greater magnitude because of the simple fact that the possibility existed that there was not a way to accomplish what I wanted to do. It was never clear at the outset how long anything could take, so the only approach was always to dive into the challenges head first and continue until they were finished.

The other issue was time management. I think back on this project and wonder what I could have done differently to get more completed. However, I cannot claim laziness or appeal to a lack of resolve. I can, however, blame a sometimes obsessive approach to getting a given task done before moving onto something else. When I destroyed a piece of equipment or had to order a new part, there were always several days of down-time. In that period of waiting, I searched for other tasks to accomplish.

## 6.2 Advantages

The advantage of this system is above all its simplicity and extensibility. Upon testing this system with a real solar power system, I realized that in some cases the essential function of alerting system overseers when the battery bank is drawn too low may actually be all that is required or practical. After all, this can be accomplished at almost no cost at all after installation. Whenever a problem arises, an alert fires, and assistance is called for. A solar system can function without monitoring if absolutely necessary, but it cannot operate if it is being progressively compromised by unintended misuse.

## 6.3 Disadvantages

There are some obvious disadvantages to this approach. First of all, there is an unfortunate amount of hardware coupling going on here. Despite the fact that the software implementation of the Modbus protocol is entirely separate from the main control software, the fact that I had to solder a wire from the VS+ output of the Max232 chip into the unused DSR (Data Signal Ready) pin on the serial port of the Freeduino unfortunately ties this implementation specifically to the TriStar-45. If another controller did not use a port-powered circuit or made handshaking use of the DSR pin, this hardware specification will be incompatible.

*6.4 Future Work*

This project has some obvious areas that remain to be implemented. The first and most obvious of these is of course the software application on the server-side of this monitoring system. This software system is for most practical purposes independent of the hardware side of this system, which is completely unaware of what party is calling it for information.

The general purpose of system is to store historical data about a solar power system, and present it in such a way that allows for long term system analysis to be conducted. Such a system will accomplish this task by periodically calling the monitoring device, collecting the data received in the form of DTMF, decoding these tones, and storing them in a database. This database can then be queried in order to produce graphs and other output representing such things as charge curves, solar insolation data, average power output over time, and so on. If a problem is detected, this system might opt to send an SMS message or other notification to one or many system overseers, but this notification has the obvious potential to be delayed heavily if a scheduler only prompts an update on the hour or every few hours. The following diagram from Appendix A shows what the theoretical output of such a system might look like. This is a manually created graph constructed from data taken during a week-long test of this system with a functional solar power system.

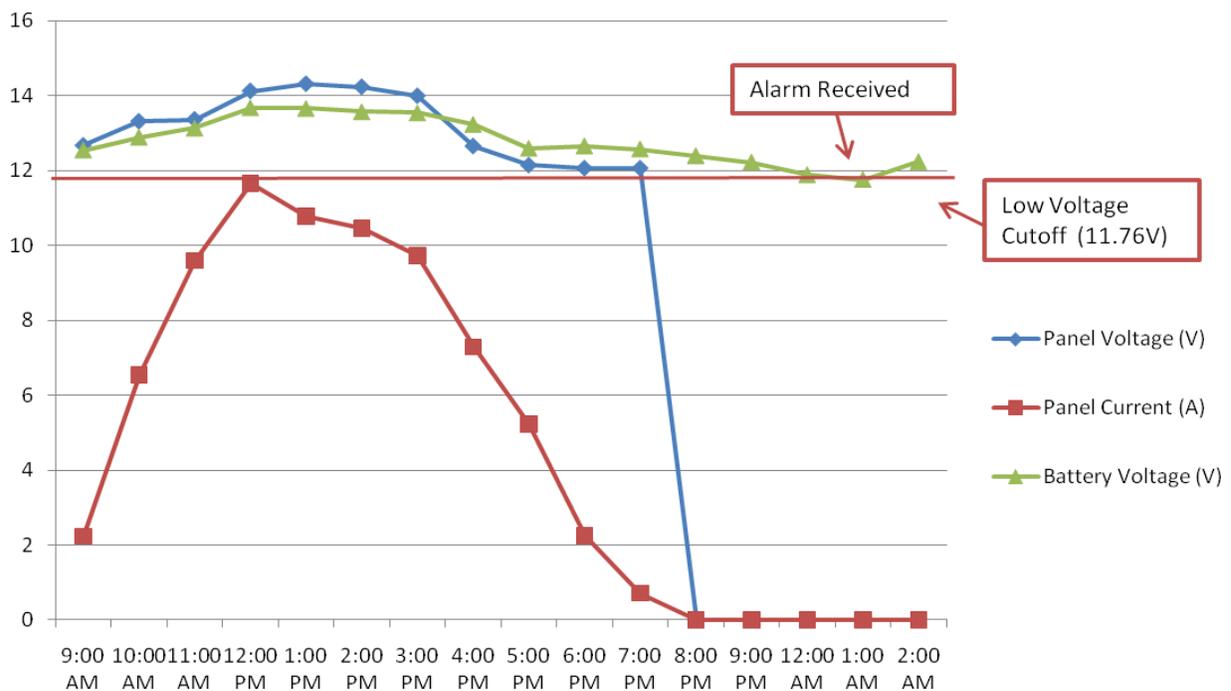

Fig 4: Manually Created Graph from Actual Measurements

The general rule, going forward, must be that if this application is to be developed and expanded upon, implementations must be carefully abstracted so as to make as few implementation-specific design choices as possible. Allowing individuals to customize applications without rewriting this project from scratch is a powerful concept, but of course this must walk a fine line between useful design assumptions and the potential confusion that may result from leaving too many critical details up to an implementation. As with any framework, sometimes making a project easily extensible for the majority of potential users is preferable to trying to accommodate every imaginable scenario.

## Acknowledgements

Thanks to Dave Small, my project advisor, who believed in my abilities and helped me with so many great insights along with way. Thanks to Dr. Gugel, who helped me with some of my more technical hardware questions despite my not even being his student this semester. Thanks to all of my friends and colleagues who entertained my ideas and contributed their thoughts to this project. Notably among them is my friend Jose Morales, who had some extraordinarily good advice at some critical moments. And of course, thanks to my family, for supporting me in every way possible during the course of this project and believing that I could do it even when I did not.

# References

1. *8-bit AVR Microcontroller with 4/8/16/32K Bytes In-System Programmable Flash*, Atmel Corporation, Rev. 8161C–AVR–05/09, (2009)

2. Anon. "Build a MaxSerial Freeduino", http://spiffie.org/electronics/archives/microcontrollers/Build%20a%20MaxSerial%20Freeduino.html, Spiff's Electronics Notebook, (as-of 9 Apr 2008)

3. Anon. "Guide to Getting Started with Arduino", http://arduino.cc/en/Guide/HomePage, Arduino (as-of 17 Nov 09)

4. Anon. "Language Reference (extended)", http://arduino.cc/en/Reference/Extended, Arduino, (as-of 17 Nov 09)

5. Donner, Jonathan. "Micro-entrepreneurs and Mobiles: An Exploration of the Uses of Mobile Phones by Small Business Owners in Rwanda." *Information Technologies and International Development,* Vol. 2, issue 1, Fall 2004.

6. Evans, Brian  W. *Arduino Programming Notebook,* Second Edition, (2008)

7. Gustavsson, Mathias, Ellegard, Anders. "The impact of solar home systems on rural livelihoods. Experiences from the Nyimba Energy Service Company in Zambia." *Renewable Energy*, Vol. 29, 2004.

8. Hankins, Mark, Van der Plas, Robert J. "Solar Electricity in Africa: A Reality." *Energy Policy*, Vol. 26, 1998.

9. Holland, Ray. "Appropriate Technology: Rural Electrification in Developing Countries." *Intermediate Technology Development Group*, Institute of Electrical and Electronics Engineers (IEEE) Review, July/August, 1989.

10. Jackson, Tim, Nhete, Tinashe, Mulugetta, Yacob. "Photovoltaics in Zimbabwe:  lessons from the GEF Solar project." *Energy Policy*, Vol. 28, 2000.

11. Karekezi, Stephen. Kithyoma, Waeni. "Renewable energy strategies for rural Africa: is a PV-led renewable energy strategy the right approach for providing modern energy to the rural poor of sub-Saharan Africa?" *Energy Policy*, Vol. 30, 2002.

12. Krause, Martin, Nordstrom, Sara. "Solar Photovoltaics in Africa: Experiences with Financing and Delivery Models." *United Nations Development Program Global Environment Facility (UNDP-GEF): Monitoring & Evaluation Report Series,* Issue 2, May 2004.

13. Lorenzo, E. "Photovoltaic Rural Electrification." *Progress in Photovoltaics: Research and Applications*, Vol. 5, 1997.

14. *MODBUS Application Protocol Specification,* Modbus-IDA, V1.1b, (2006)

15. *Modicon Modbus Protocol Reference Guide*, MODICON Inc. Industrial Automation Systems**,** PI–MBUS–300 Rev. J, (1996)

16. *Optically Coupled Isolator: OPI110, OPI113, OPI1264 Series*, OPTEK Technology Inc., Issue A.1, (2007)

17. Togola, Ibrahim. *Renewable Energy Solution Perspectives for Africa.* Mali Folkcenter, Montreal, December 1, 2005.

18. *TriStar MODBUS Specification*, Morningstar Corporation, Version 2, (2005)

19. *TriStar Solar System Controller Installation and Operation Manual,* Morningstar Corporation (2009)

20. Wamukonya, Njeri. "Socioeconomic Impacts of Rural Electrification in Namibia: Comparisons Between Grid, Solar and Unelectrified Households." *Energy for Sustainable Development*, United Nations Environment Program Collaborating Centre on Energy and Environment (UNEP-UCCEE), Volume 5, No. 3, September 2001.

## Biography

My name is Nikolas Wolfe. I live on the planet Earth. I was born there, and have yet to leave. I go to school at the University of Florida, and provided things work out with this project (wink wink!) I'll escape in a week or so with a degree.